# Variation of Strong Correlation Effects in A-site Ordered Perovskites $CaCu_3Ti_{4-x}Ru_xO_{12}$: Photoemission and Inverse Photoemission Studies

Hojun Im[1*], Masataka Iwataki[1], Masanori Tsunekawa[2], Takao Watanabe[1], Hitoshi Sato[3], Masashi Nakatake[3,4], Shin-ichi Kimura[5,6]

[1] *Graduate School of Science and Technology, Hirosaki University, Hirosaki 036-8224, Japan*

[2] *Faculty of Education, Shiga University, Otsu 520-0862, Japan*

[3] *Hiroshima Synchrotron Radiation Center, Hiroshima University, Higashi-Hiroshima 739-0046, Japan*

[4] *Aichi Synchrotron Radiation Center, Aichi Science and Technology Foundation, Seto 489-0965, Japan*

[5] *Graduate School of Frontier Biosciences, Osaka University, Suita 565-0871, Japan*

[6] *UVSOR Facility, Institute for Molecular Science, Okazaki 444-8585, Japan*

**Abstract**

We have systematically studied the strong correlation effects in A-site ordered perovskites $CaCu_3Ti_{4-x}Ru_xO_{12}$ (x = 0, 1, 3.5, and 4) by using photoemission and inverse photoemission spectroscopies. In x = 0, 1, 3.5, the peak positions of the strongly correlated Cu 3d states around -3.8 eV and Ti 3d states around 3.6 eV little change. On the other hand, in x = 4, the Cu 3d states split into two peaks around -2.5 and -4 eV. These indicate that Ti plays an important role to retain the strong correlation effects. In addition, the multiplet structures of Cu 3d final states from -8 to -14 eV become weak as Ru increases, indicating the reduction of the localized characters of Cu 3d states. At the Fermi level, we observe the absence of spectral weight in x = 0, 1 and the development of Ru 4d in-gap states in x = 3.5, 4, which give rise to the metal-insulator transition between x = 1 and x = 3.5.

## 1. Introduction

Strong correlation effects have been long-standing issues in condensed matter physics due to the intriguing phenomena, such as Mott-insulator, heavy fermion, high-Tc superconductivity and quantum critical phenomena [1)–5)]. Among them, A-site ordered perovskites, $CaCu_3Ti_4O_{12}$ (CCTO) and $CaCu_3Ru_4O_{12}$ (CCRO), are one of the most widely studied systems. CCTO has been well known as the Mott-insulator with extremely high dielectric constant [6)–8)]. The origin of the Mott insulating state in CCTO has been investigated in terms of the electronic structure by a variety of experimental and theoretical methods [9)–14)]. In the photoemission (PES) and inverse photoemission (IPES) measurements, we have revealed that Cu 3d and Ti 3d electrons hybridized with O 2p states are strongly correlated, causing the absence of the density of states near the Fermi level ($E_F$) and the band shift of Cu 3d and Ti 3d peaks away from each other in comparison with the local-density approximation (LDA) calculations [11)–13)]. CCRO has attracted much attention due to heavy-fermion like behaviors [15)–17)], which have been usually observed in strongly correlated 4f-electron systems [4),18)]. It has been widely accepted that the localized Cu 3d electrons and the itinerant Ru 4d electrons cause the heavy-fermion behaviors via Kondo effects [15)]. These electronic structures have been also explained by considering the correlation effects in photoemission and band calculation studies [19)–22)]. In addition, electrical transport measurements on $CaCu_3Ti_{4-x}Ru_xO_{12}$ have revealed that the metal-insulator transition takes place between x = 1.5 and 4 [15),23)]. It is certain that $CaCu_3Ti_{4-x}Ru_xO_{12}$ is a proper system to clarify strong correlation effects in d-electron systems. However, systematic studies of their electronic structure have been insufficient to explain the mechanism of the above intriguing phenomena. To this end, we have systematically studied the electronic structure of $CaCu_3Ti_{4-x}Ru_xO_{12}$ (x = 0, 1, 3.5, 4) in both the occupied and unoccupied energy regions by the PES and IPES spectroscopies, respectively.

## 2. Experimental Details

We have synthesized the polycrystalline samples of $CaCu_3Ti_{4-x}Ru_xO_{12}$ (x = 0, 1, 3.5, 4) by a conventional solid-state reaction method. Their single phase was confirmed by X-ray diffraction pattern. The PES measurements have been carried out at the beamline BL-7 of HiSOR [24] after pilot experiments at the beamline BL5U of UVSOR. The used photon energies (*hv*) were 40 and 70 eV. The PES spectra were obtained at T = 300 K for x = 0, 1, 3.5 and T = 100 K for x = 4. The energy resolution was set to about 35 meV at *hv* = 40 eV. The IPES experiments have been performed by a stand-alone apparatus with the tunable photon energy mode at HiSOR [25),26)]. The IPES spectra (x = 1, 3.5, 4) were obtained at the kinetic energies of incident electrons ($E_k$) of 40, 46, 50 eV and T = 300 K. The energy resolution was estimated to be about 650 meV at $E_k$ = 40 eV. The clean surfaces of samples were prepared by *in situ* cleaving in the ultra-high vacuum. The Fermi level was referred to that of Au. For the IPES spectra of x = 0, we refer to the results obtained in our previous studies [12)].

## 3. Results and Discussion

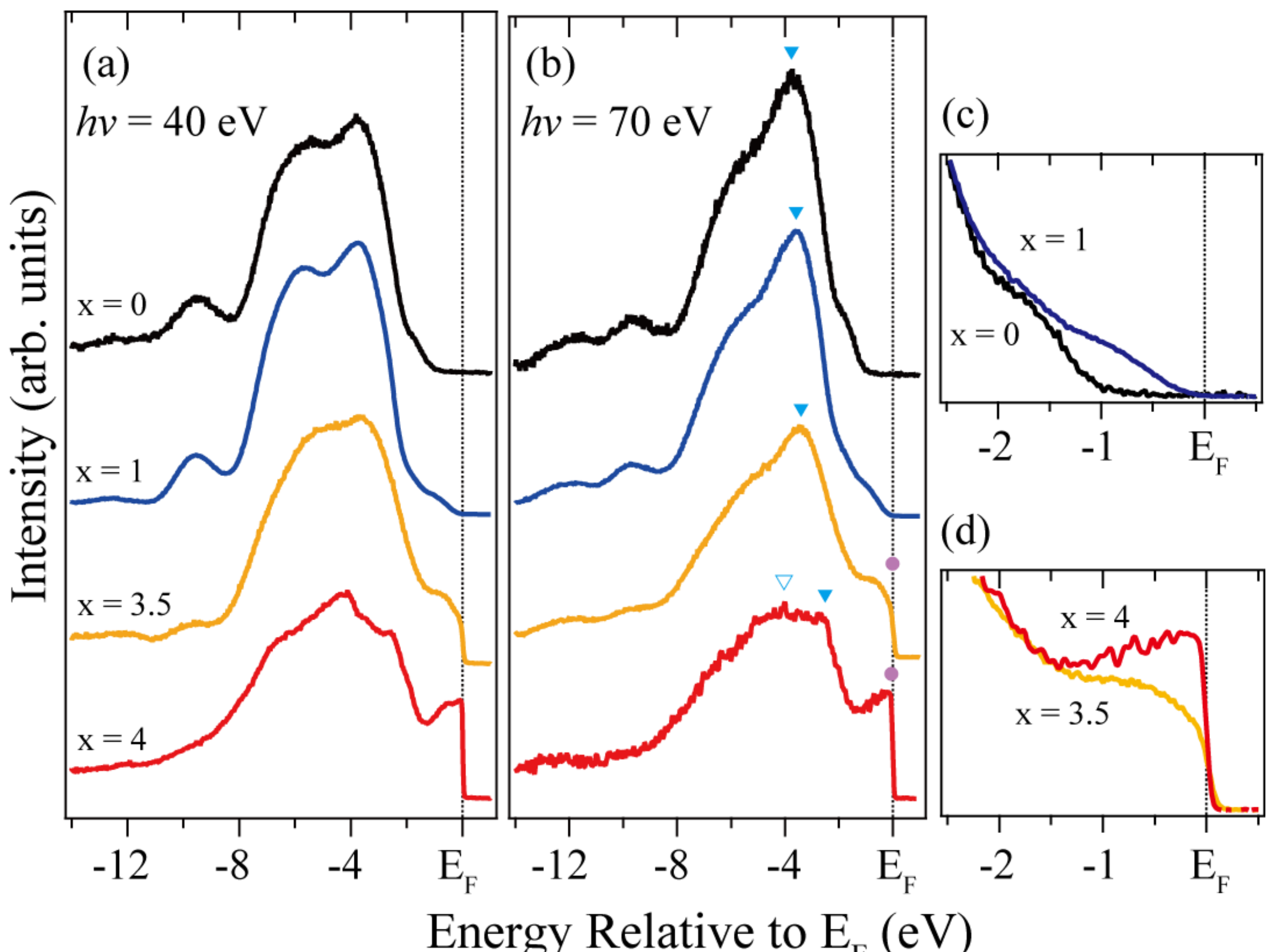

Fig. 1. (Color online) (a, b) PES spectra of $CaCu_3Ti_{4-x}Ru_xO_{12}$ obtained at *hv* = 40 and 70 eV in the valence band region. The peak positions of the Cu 3d and Ru 4d states were denoted by the triangles and solid circles, respectively. (c, d) PES spectra of $CaCu_3Ti_{4-x}Ru_xO_{12}$ near the Fermi level for the insulating (x = 0, 1) and metallic (x = 3.5, 4) states at *hv* = 70 eV. The PES spectra of x = 0 were shifted toward $E_F$ by 0.22 eV. Measurement temperatures were 300 K for x = 0, 1, 3.5 and 100 K for x = 4.

Figure 1 shows the PES spectra of $CaCu_3Ti_{4-x}Ru_xO_{12}$ (x = 0, 1, 3.5, 4) in the valence band region and near $E_F$. They were measured at *hv* = 40, 70 eV and T = 300 K for x = 0, 1, 3.5 and T = 100 K for x =4. The different measurement temperatures do little affect the feature of the PES spectra in the valence band region, except for the thermal broadening of the Fermi edge, e.g. about 100 meV at room temperature, which can be ignored for the sake of argument in this work. All PES spectra of x = 0 were shifted toward $E_F$ by 0.22 eV to compensate the charge-up effects, which have sometimes been observed due to the good insulating state of CCTO and the strong intensity of the incident light [12),13)]. Let us firstly discuss the feature of the obtained PES spectra in CCTO (x = 0), comparing with our previous works [11)–13)]. We observe that the spectral weights are negligible at $E_F$, revealing the insulating state. There are the intensive peaks around -3.8 eV of mainly Cu 3d states and the hump around -6 eV of mainly O 2p states. In the region from -8 to -14 eV, we observe the small double peaks, which come from the multiplet structures of Cu 3d final states ($d^8$ and $d^{10}\underline{L}^2$ around -9 eV, and mainly $d^8$ final states around -12 eV, where $\underline{L}$ is a ligand with a hole) as in CuO [27)]. Recent X-ray Raman scattering experiments also support the similarity between CCTO and CuO, revealing the divalent of Cu 3d states [14)]. These features of the PES spectrum have been well understood on the base of the strong correlation effects [12),28)]. For the further details, refer to our previous papers [11)–13)]. Here, we would like to focus on the variation of Cu 3d peak intensity around -3.8 eV, which increases as *hν* increases from 40 to 70 eV due to the increase of photoionization cross section of Cu 3d states [29),30)].

In the case of CCRO (x = 4), we can recognize that there are the large spectral weights near $E_F$ in contrast to CCTO, indicating the metallic state. In the region from -2 to -8 eV, there are intensive broad peaks, which seem to be superposition of three peaks around -2.5, -4, and -6 eV. In the band calculations, the spectral weights near $E_F$ have been attributed to the Ru 4d states [19)]. The Cu 3d states have contributed to the spectral weights mainly around -2.5 eV and partially around -4 eV, which has been addressed in terms of the multiplet structure of Cu 3d final states [19)]. The spectral weights around -6 eV are composed of mainly the O 2p states [19),20)]. Here, it should be noted that the Cu 3d peak around -2.5 eV increases with *hν*. In terms of the photoionization cross section, the Cu 3d states around -3.8 eV in CCTO and around -2.5 eV in CCRO have similar characters. This has been also supported by the previous reports where the peak around -2.5 eV of CCRO is mainly composed of the Cu 3d states [19),20),22)]. In CCRO, the multiplet structure of Cu 3d final states around -9 eV is not clearly observed compared to that of CCTO. This indicates that the localized character of Cu 3d states becomes weak with the substitution of Ru for Ti. Note that the localized character of Cu 3d states still exists as shown in the observation of the multiplet structure around -12 eV.

Next, let us discuss the PES spectra of $CaCu_3Ti_{4-x}Ru_xO_{12}$, considering the variation of the Ru concentration (x = 0, 1, 3.5, 4). Fig. 1(c) shows the PES spectra of x = 0 and 1 near $E_F$ at $h\nu$ = 70 eV. In x = 1, we observe the spectral weight around -1 eV in addition to the small shoulder around -1.7 eV, which is composed of Cu 3d-O 2p hybridized bands and corresponds to the lower Hubbard band in x = 0 [11)]. However, these spectral weights do not extend to $E_F$, retaining the insulating state. Except for the region near $E_F$, the spectral shapes of x = 0 and 1 are very similar in the valence band region as shown in Fig. 1(a) and 1(b). On the other hand, the spectral weights of x = 3.5 and 4 are clearly observed at $E_F$ as shown in Fig. 1(d), showing the metal-insulator transition between x = 1 and x = 3.5 in good agreement with the results of transport experiments [15),23)]. It is found that the spectral shape of x = 3.5 is similar with those of x = 0, 1 in the valence band region, except for the decrease of Cu 3d peak around -3.8 eV (Fig. 1(a) and 1(b)). Here, we should note that the peak positions of Cu 3d state around -3.8 eV little change in x = 0, 1, 3.5. On the other hand, the Cu 3d peaks split into two peaks around -2.5 and -4 eV in CCRO (x = 4) where there is no Ti concentration. Even a small Ti concentration seems to pin the position of the Cu 3d states around -3.8 eV. This indicates that the Ti 3d states play an important role to retain the strong correlation effects in the $CaCu_3Ti_{4-x}Ru_xO_{12}$ system. In addition, we would like to emphasize that the intensity of the multiplet structure of the Cu 3d final states in the region from -8 to -14 eV decreases with increasing x. This means that the increase of Ru concentration reduces the atomic-like characters of Cu 3d states caused by the strong correlation effects.

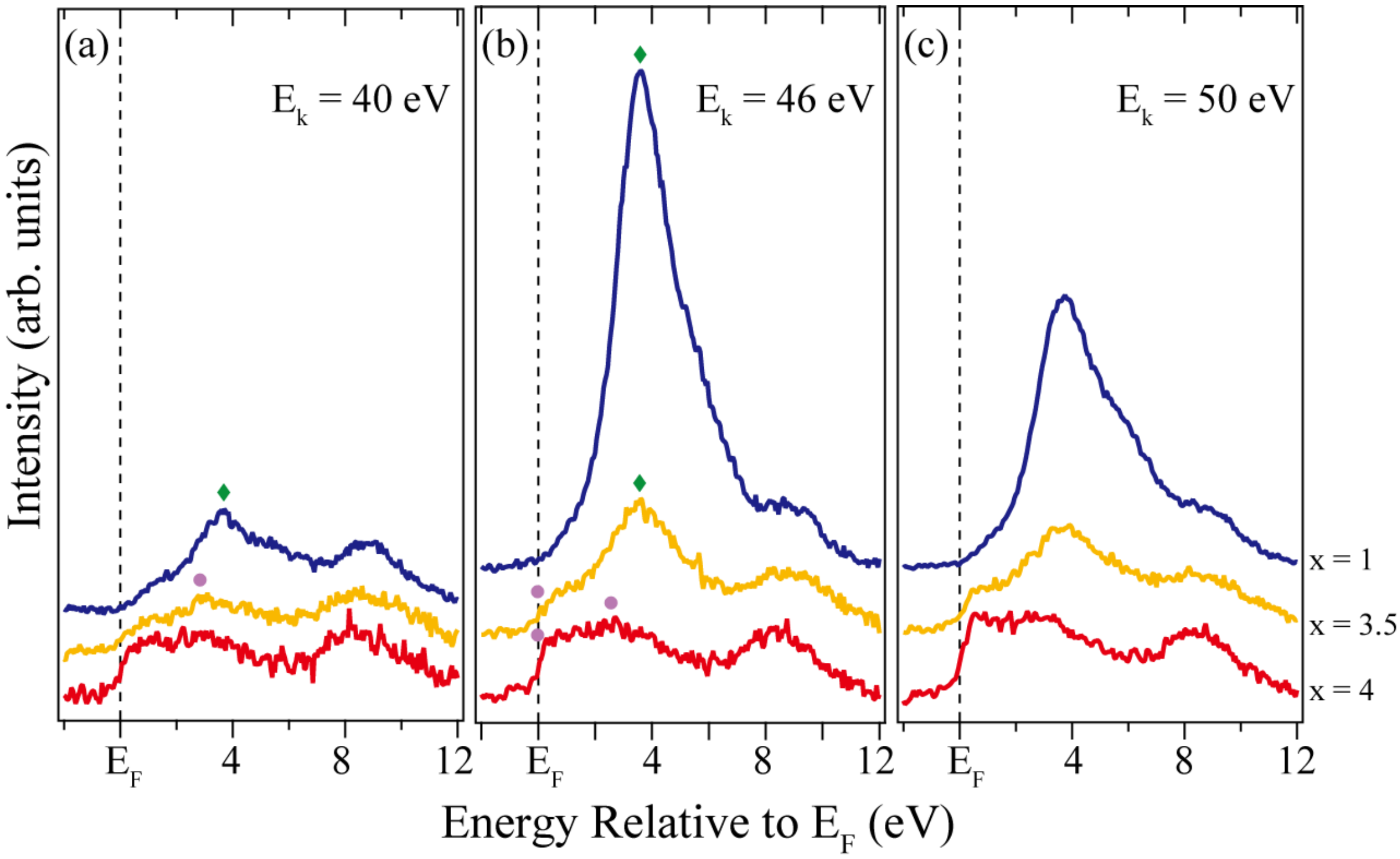

Fig. 2. (Color online) IPES spectra of $CaCu_3Ti_{4-x}Ru_xO_{12}$ obtained at $E_k$ = 40, 46, and 50 eV in the valence band region. The solid diamonds and solid circles denote the positions of the Ti 3d and Ru 4d states, respectively. All spectra were measured at T = 300 K.

Figure 2 shows the IPES spectra of $CaCu_3Ti_{4-x}Ru_xO_{12}$ (x = 1, 3.5, 4) in the unoccupied energy region from 0 to 12 eV. They were obtained at $E_k$ = 40, 46, 50 eV and T = 300 K, and were normalized to the intensity of the Ca 3d peak around 9 eV [12]. In the case of x = 1, we observe the negligible spectral weights near $E_F$, indicating the insulating state in agreement with the PES results. We clearly observe that the Ti 3d peak around 3.6 eV is largely enhanced at $E_k$ = 46 eV as shown in Fig. 2(b), due to the Fano-resonance between standard IPES process and Coster-Kronig Auger process around Ti 3p-3d edge [30),31)]. The resonant effects are much obvious in comparison with that of CCTO (x = 0) in our previous paper [12]. This is possibly because the substitution of Ru increases the conductivity of sample, which reduces the charge-up of the sample, and hence make measurement easier. Further studies are required for clear comprehension. In CCRO (x = 4), the IPES spectrum does not show the resonant effects due to the absence of the Ti concentration. Instead of that, the spectral weight of Ru 4d states near $E_F$ increases with $E_k$ due to the change of the photon-energy dependent cross section of Ru 4d state, whose matrix element of IPES is similar to that of PES in the relation of the time reversal [32),33)]. This reveals that the Ru 4d states have an itinerant character. In Figs. 2(b) and 2(c) where the IPES spectra were obtained at $E_k$ = 46 and 50 eV, respectively, the peak positions of Ti 3d states of x = 1 and 3.5 are almost the same as that of CCTO (x = 0) in our previous paper [12)], where the Ti 3d peak is located in the higher energy region in comparison with the LDA calculations due to the strong correlation effects. On the other hand, in Fig. 2(a), the IPES spectra were obtained at $E_k$ = 40 eV which is close to an off-resonant condition of the Ti 3p-3d edge [31)]. There is a peak around 3 eV in x = 3.5, instead of the Ti 3d peak around 3.6 eV. This peak can be assigned to Ru 4d states as predicted in the band calculations [19)]. The spectral weight of Ru 4d states is also observed at $E_F$ in x = 3.5, indicating metallic state, in agreement with the PES results.

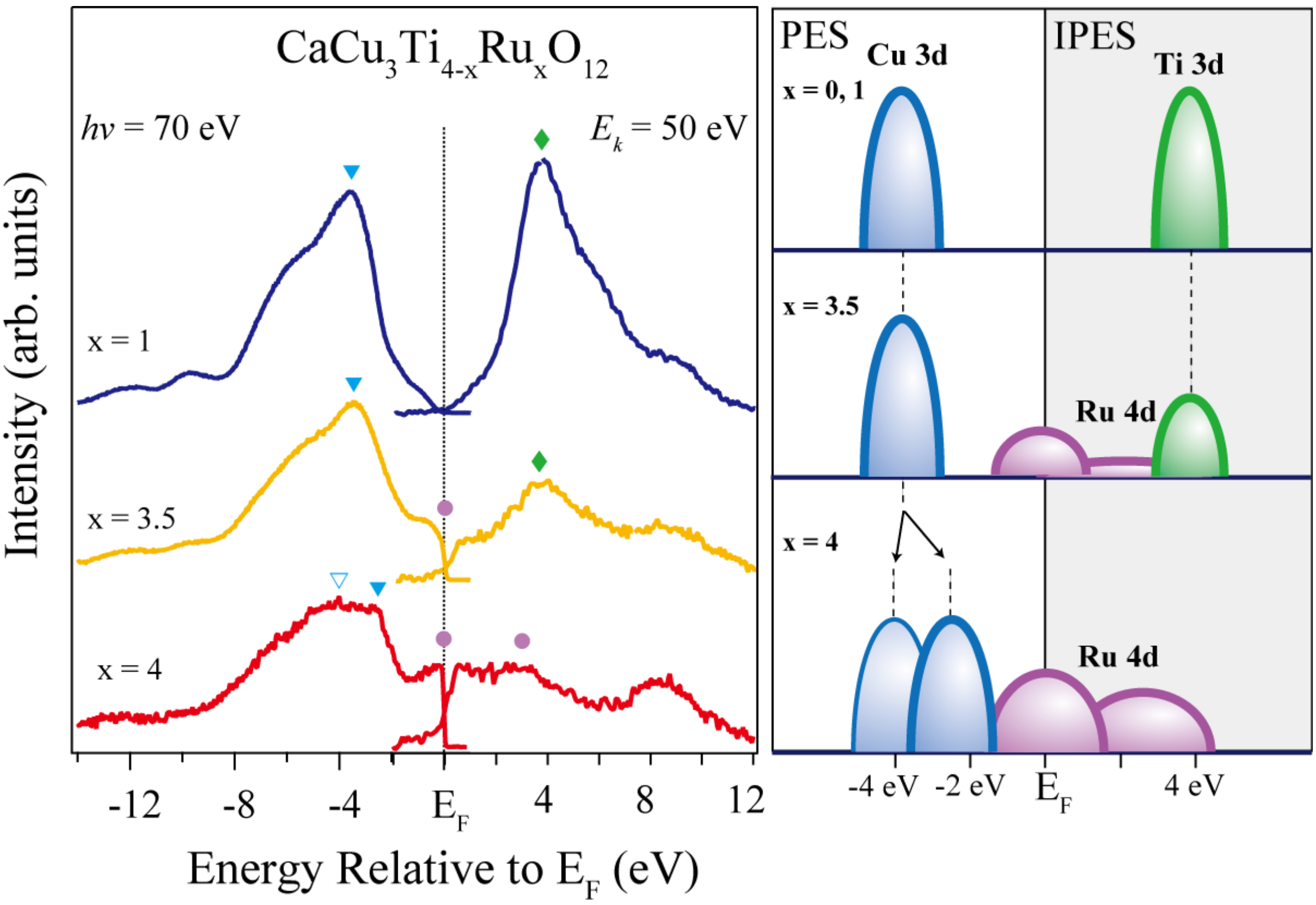

Fig. 3. (Color online) (Left panel) PES and IPES spectra of $CaCu_3Ti_{4-x}Ru_xO_{12}$ (x = 1, 3.5, 4) obtained at *hν* = 70 eV and $E_k$ = 50 eV, respectively. The triangles, solid diamonds, and solid circles denote the positions of the Cu 3d, Ti 3d, and Ru 4d states, respectively. (Right panel) Schematic diagram of the spectral weight variation of the Cu 3d, Ti 3d, and Ru 4d orbitals with respect to the Ru concentration.

Finally, let us discuss the metal-insulator transition in $CaCu_3Ti_{4-x}Ru_xO_{12}$, focusing on the d-states in both the occupied and unoccupied energy regions. The left panel of Fig. 3 shows the PES and IPES spectra at *hν* = 70 eV and $E_k$ = 50 eV, respectively. A corresponding schematic diagrams of spectral weights of the d-orbitals are shown in the right panel of Fig. 3. Both peak positions of Cu 3d states around -3.8 eV and Ti 3d states around 3.6 eV little change in x = 1, 3.5, while the Cu 3d states split into two peaks around -2.5 and -4 eV in x = 4. These emphasize the important role of the Ti 3d states in the strong correlation effects. As the Ru concentration increases, the in-gap states of Ru 4d states were developed at $E_F$ in x = 3.5 and more clearly observed in x = 4. Consequently, the metal-insulator transition has been attributed to the appearance of the Ru 4d in-gap states at $E_F$ between the peaks of the strongly correlated Cu 3d and Ti 3d states rather than the variation of the ratio between bandwidth and on-site Coulomb energy. Such behaviors are different with the mechanism of the conventional Mott metal-insulator transition [1),2)]. These results also mean that the localized characters of Cu 3d states and the itinerant characters of Ru 4d states coexist, supporting the understanding of the heavy fermion behaviors of CCRO (x = 4), which have originated from the Kondo effects formed by the Cu 3d and Ru 4d electrons [15)]. For further understanding, the obtained PES and IPES results should be compared with theoretical studies such as a dynamical mean-field theory.

## 4. Conclusion

To systematically study the variation of strong correlation effects in A-site ordered perovskites $CaCu_3Ti_{4-x}Ru_xO_{12}$, we have performed the PES and IPES experiments. The peak positions of Cu 3d states around -3.8 eV and Ti 3d states around 3.6 eV little change in x = 0, 1, 3.5, while Cu 3d states split into two peaks around -2.5 and -4 eV in x = 4. In addition, the intensity of the multiplet structures of Cu 3d final states in the region from -8 to -14 eV becomes small as the Ru concentration increases. These results have revealed that the strong correlation effects become weak due to the substitution of Ru for Ti. We also found that the absence of spectral weight at $E_F$ in x = 0, 1 and the Ru 4d in-gap states at $E_F$ in x = 3.5, 4 have been responsible for the metal-insulator transition.

**Acknowledgments** This work was partly supported by the HiSOR Facility Program (14-A-60 and 14-A-61) of Hiroshima University and by the UVSOR Facility Program (25-806) of the Institute for Molecular Science.

*hojun@hirosaki-u.ac.jp

1) M. Imada, A. Fujimori and Y. Tokura, Rev Mod Phys **70** [4], 1039 (1998).

2) N. Mott, *Metal-Insulator Transitions* (Taylor & Francis, London, 1974).

3) J. G. Bednorz and K. A. Müller, Z. Physik B - Condensed Matter **64** [2], 189 (1986).

4) H. J. Im, T. Ito, H.-D. Kim, S. Kimura, K. E. Lee, J. B. Hong, Y. S. Kwon, A. Yasui and H. Yamagami, Phys Rev Lett **100** [17], 176402 (2008).

5) H. J. Im, T. Ito, J. B. Hong, S. Kimura and Y. S. Kwon, Phys Rev B **72** [22], 220405 (2005).

6) M. A. Subramanian, D. Li, N. Duan, B. A. Reisner and A. W. Sleight, J Solid State Chem **151** [2], 323 (2000).

7) A. P. Ramirez, M. A. Subramanian, M. Gardel, G. Blumberg, D. Li, T. Vogt and S. M. Shapiro, Solid State Commun **115** [5], 217 (2000).

8) C. C. Homes, T. Vogt, S. M. Shapiro, S. Wakimoto and A. P. Ramirez, Science **293** [5530], 673 (2001).

9) L. He, J. B. Neaton, M. H. Cohen, D. Vanderbilt and C. C. Homes, Phys Rev B **65** [21], 214112 (2002).

10) Y. Zhu, J. C. Zheng, L. Wu, A. I. Frenkel, J. Hanson, P. Northrup and W. Ku, Phys Rev Lett **99** [3], 037602 (2007).

11) H. J. Im, M. Tsunekawa, T. Sakurada, M. Iwataki, K. Kawata, T. Watanabe, K. Takegahara, H. Miyazaki, M. Matsunami, T. Hajiri and S. Kimura, Phys Rev B **88** [20], 205133 (2013).

12) H. J. Im, M. Iwataki, S. Yamazaki, T. Usui, S. Adachi, M. Tsunekawa, T. Watanabe, K. Takegahara, S. Kimura, M. Matsunami, H. Sato, H. Namatame and M. Taniguchi, Solid State Commun **217**, 17 (2015).

13) H. J. Im, T. Sakurada, M. Tsunekawa, T. Watanabe, H. Miyazaki and S. Kimura, Solid State Commun **298**, 113648 (2019).

14) Y. Tezuka, Y. Yokouchi, S. Sasaki, S. Nakamoto, K. Nishiyama, M. Mikami, H. J. Im, T. Watanabe, S. Nozawa, N. Nakajima and T. Iwazumi, J Electron Spectros Relat Phenomena **220**, 114 (2017).

15) W. Kobayashi, I. Terasaki, J. Takeya, I. Tsukada and Y. Ando, J Physical Soc Japan **73** [9], 2373 (2004).

16) A. Krimmel, A. Günther, W. Kraetschmer, H. Dekinger, N. Büttgen, A. Loidl, S. G. Ebbinghaus, E.-W. Scheidt and W. Scherer, Phys Rev B **78** [16], 165126 (2008).

17) S. Tanaka, N. Shimazui, H. Takatsu, S. Yonezawa and Y. Maeno, J Physical Soc Japan **78** [2], 024706 (2009).

18) H. J. Im, T. Ito, H. Miyazaki, S. Kimura, Y. S. Kwon, Y. Saitoh, S.-I. Fujimori, A. Yasui and H. Yamagami, Solid State Commun **209–210**, 45 (2015).

19) N. Hollmann, Z. Hu, A. Maignan, A. Günther, L.-Y. Jang, A. Tanaka, H.-J. Lin, C. T. Chen, P. Thalmeier and L. H. Tjeng, Phys Rev B **87** [15], 155122 (2013).

20) T. T. Tran, K. Takubo, T. Mizokawa, W. Kobayashi and I. Terasaki, Phys Rev B **73** [19], 193105 (2006).

21) H. Xiang, X. Liu, E. Zhao, J. Meng and Z. Wu, Phys Rev B **76** [15], 155103 (2007).

22) T. Sudayama, Y. Wakisaka, K. Takubo, T. Mizokawa, W. Kobayashi, I. Terasaki, S. Tanaka, Y. Maeno, M. Arita, H. Namatame and M. Taniguchi, Phys Rev B **80** [7], 075113 (2009).

23) A. P. Ramirez, G. Lawes, D. Li and M. A. Subramanian, Solid State Commun **131** [3–4], 251 (2004).

24) M. Taniguchi and J. Ghijsen, International Union of Crystallography Journal of Synchrotron Radiation J. Synchrotron Rad **5**, 1176 (1998).

25) H. Sato, T. Kotsugi, S. Senba, H. Namatame, M. Taniguchi and B. Hiroshima, International Union of Crystallography Journal of Synchrotron Radiation J. Synchrotron Rad **5**, 772 (1998).

26) H. Sato, M. Arita, Y. Utsumi, Y. Mukaegawa, M. Sasaki, A. Ohnishi, M. Kitaura, H. Namatame and M. Taniguchi, Phys Rev B **89** [15], 155137 (2014).

27) J. Ghijsen, L. H. Tjeng, J. van Elp, H. Eskes, J. Westerink, G. A. Sawatzky and M. T. Czyzyk, Phys Rev B **38** [16], 11322 (1988).

28) H. Eskes, L. H. Tjeng and G. A. Sawatzky, Phys Rev B **41** [1], 288 (1990).

29) J. J. Yeh and I. Lindau, At Data Nucl Data Tables **32** [1], 1 (1985).

30) S. Hüfner, *Photoelectron Spectroscopy* (Springer-Verlag, Berlin, 1995).

31) M. Arita, H. Sato, M. Higashi, K. Yoshikawa, K. Shimada, M. Nakatake, Y. Ueda, H. Namatame, M. Taniguchi, M. Tsubota, F. Iga and T. Takabatake, Phys Rev B **75** [20], 205124 (2007).

32) Th. Fauster and F. J. Himpsel, Phys Rev B **30** [4], 1874 (1984).

33) J. B. Pendry, Phys Rev Lett **45** [16], 1356 (1980).